\documentclass[preprint,aps,pre,amsmath,amssymb]{revtex4}
\usepackage{graphics}
\usepackage{epsf,graphicx,pstcol,epsfig,psfrag}
\begin{document}

\title{Ising spin ladder with trimer rungs and next-nearest-neighbor coupling. Frustration in physics and agent models.}
\author{Jozef Sznajd}
\affiliation{Institute for Low Temperature and Structure Research, Polish Academy of Sciences, 50-422 Wroclaw, Poland} 
\date{\today}

\begin{abstract}
The extended model of two-leg Ising spin ladder with trimer rungs and  next nearest neighbor interaction (NNN) in an external magnetic field 
is studied using the transfer matrix and  linear renormalization group methods. In the standard version (with the same only NN interactions in both legs) such a ladder exhibits very interesting behavior - a frustration driven extremely sharp phase crossover at finite temperature which resembles a phase transition, impossible in one dimension. 
 It is shown that in all considered cases with different interactions in each leg (assymetric ladder), with NNN interactions and in the presence of small external field such a crossover takes place when the point  at which the effective interleg coupling vanishes coincides with the point at which the interleg correlation function exhibits an inflection point  accompanied by the specific heat  maximum.
 
 A hypothesis is formulated  that to describe an abrupt change in political view of the people it is not necessary to resort to a concept of a phase transition, controversial for social systems. 
In some cases, this phenomenon can be understood as a phase crossover triggered by the level of frustration.
The source of this frustration is proposed as an internal conflict between two areas of attitudes of the society members, personal and economic, which are shaped by personality traits possibly modified by life experience.

\end{abstract}
\maketitle

\section{Introduction}

It has been shown in numerous papers \cite{Gal, Str, Souza, Rojas, Wei1, Wei2, Hut, Tsv, Rojas2} that some one-dimensional spin models exhibit a peculiar  behavior, namely they show discontinuities or very sharp (gigantic) maxima in thermodynamic quantities at finite temperature. This behavior, named marginal phase transition or pseudo-transition or simply phase crossover at finite temperature has been observed  in decorated spin systems, where after removing redundant spins the effective interaction between remaining spins vanishes at some temperature.
The paradigm  is a spin ladder in the presence of an additional frustrating rung spins. In such systems, at a certain temperature $T=T_p$ one can observe a decoupling of the two legs of the ladder  i.e. vanishing of the effective interleg interaction due to the competition between the interleg and rung couplings. 
One of such systems is
the two-leg Ising spin ladder with trimer rungs (Fig.1) which attracts special interest because of its simplicity \cite{ Wei1, Wei2, Hut, Tsv, Krok}. The model described by the Hamiltonian 
\begin{eqnarray}
H=-\tilde{J}\sum_{i}(S_i S_{i+1}+V_i V_{i+1})-\tilde{J_1} \sum_{i}B_i ( S_i +V_i )-\tilde{J_2} \sum_{i}S_i V_i.
\end{eqnarray}
was solved exactly by the transfer matrix method \cite{Wei1}. The free energy of the model (1) has been evaluated in two steps \cite{Wei2, Hut}: first,  partial summation over $B_i$  spins degrees of freedom was performed in a rigorous way, which leads to an effective Hamiltonian with a spin-independent term $A$  and  a temperature dependent effective interaction between $S_i$ and $V_i$ spins. The transfer matrix method was then applied to the standard two-leg ladder. So the Hamiltonian of the two-leg Ising spin ladder with trimer rungs after summing over $B_i$ spins can be rewritten in the form
\begin{eqnarray}
H_r= N A+j\sum_{i=1}(S_i S_{i+1}+V_i V_{i+1})+j_p  \sum_i  S_i V_i
\end{eqnarray}
where the factor $-1/T$ has been absorbed in the Hamiltonian ($j=-\frac{\tilde{J}}{T}$)
and as was shown by Weiguo Yin \cite{Wei1}
\begin{equation}
A= \frac{1}{2}[\log 2 \cosh \frac{ 2 j_1}{T}+\log 2],  \quad j_p= -\frac{\tilde{J_2}}{T} + B T, \quad B= \frac{1}{2}[\log 2 \cosh \frac{ 2 j_1}{T}-\log 2 ],
\end{equation}
and $j_1 = -\frac{\tilde J_1}{T}$.

The results for the entropy, specific heat  and correlation length of such a model were recently discussed in detail in Ref. \cite{Hut}.
The authors concluded that the observed anomalies, jump of the entropy, sharp peak of the specific heat  and the extremely large  correlation length around the temperature $T=T_p$ at which the effective interleg interaction vanishes $(j_p=0)$ are a reminiscence of the standard criticality ot the Ising chain  at zero temperature. At $T=T_p$ the two ladder legs are decoupled and  two standard Ising chains are obtained.  They also framed a conjecture that to exhibit discussed anomalies
\emph{"frustration is not vitally necessary; the only demand is to have a suitable}  $j_p(T)$".
The conclusion is general and applies to a whole class of systems exhibiting similar behavior. So the question can be asked whether similar anomalies are observed in an asymmetric ladder with different interaction parameters in the legs and next-nearest-neighbor interactions. 

The two-leg spin ladder can be treated as a chain with two spins in each site which is related to the Hamiltonian formulation of the dynamical model simulating the process of opinion spreading in a society, where people adopt one of the four social attitudes \cite{KW, JS1}. 
The model derived from sociophysical considerations \cite{JS1} (ACLS model) contains the nearest-neighbor (NN) interaction between each species of spins $S_i$ and $V_i$,  where the first
 represents one of  two possible opinions ("Yes" or "No") in the personal area and second
 in the economic area, interaction between the two kind of species and additional the next-nearest-neihgbor (NNN)  interaction  between $V_i$ spins. The two different kind of interactions between $S_i$ spins (only NN) and $V_i$ spins (NN and NNN) refers to two kinds of the information flows in dynamic sociophysics model: inward from the neighborhood in the personal area and outward to the neighborhood  in the economic area.  It has been shown \cite{JS1} that the insite coupling ($S_i V_i$) can trigger or remove a specific heat maximum which can indicate a kind of local (short range) reordering.  
Let us assume that this coupling (mutual influence of the personal and economic areas) can be modified by 
some  personal predispositions of agents described by $\tilde J_1$ (1). Now the generalized two-leg ladder with trimer rungs, nonequivalent legs and NNN interaction can define stable states of the dynamical model describing opinion spreading in a society. The object of our interest is the role of  frustration and the next-nearest-neighbor interaction in the emergence of various behaviors of such an asymmetric ladder model. 
 
Unlike the standard two-leg ladder with trimer rungs a generalized model with NNN interactions cannot be solved exactly. Therefore, to find the thermodynamic properties of such a model, we use the linear renormalization group (LRG) transformation.
So, in Sec.II the validity of the  LRG for Ising spin ladders is discussed. In Sec. III the transfer matrix method is used to study symmetric (standard) and asymmetric ladders with NN interactions in an external magnetic field. In Sec. IV the LRG method is applied to generalized asymmetric spin ladder with NNN interactions and in Sec.V the usefulness of such a ladder for describing the stable fixed point of the dynamical socio-physical model ACLS is discussed . Finally, the results are summarized in Sec. V.

\section{Renormalization of the two-leg Ising spin ladder}

In this section we test the applicability of the renormalization group technique to describe the thermodynamic properties of the Ising spin ladder model. We apply the linear renormalization group (LRG) transformation (decimation) defined by
\begin{equation}
e^{H'_r}=Tr_{S,V} P(\sigma, \upsilon; S,V) e^{H_{r}}.
\end{equation}
The weight operator $P(\sigma, \upsilon; S,V) $ which couples the original spins $S_i, V_i$ and effective $\sigma_i, \upsilon_i$ ones  is chosen in the linear form
\begin{equation}
P(\sigma, \upsilon; S,V)=\prod_{i=0}^{N} p_i(\sigma, \upsilon; S,V)
\end{equation}
and
\begin{equation}
 p_i(\sigma, \upsilon; S,V) =\prod_{n=0}^{\omega} (1+\sigma_{i+n} S_{i+nm})(1+\upsilon_{i+n} V_{i+nm}).
\end{equation}

The next step is to choose of a renormalized block or, in other words, values of $m$ and $\omega$. It is well known that the LRG  transformation is exact for the Ising chain regardless of the block size \cite{MN}. However, in this section  we use the smallest block that allows us to consider both ferromagnetic and antiferromagnetic ground state structure i.e. a  4-rung  one (Fig.1), which means $m=3$ and $\omega =1$ and the linear RG projector has the form
\begin{equation}
 p_i(\sigma, \upsilon; S,V) = (1+\sigma_i S_i)(1+\sigma_{i+1} S_{i+3})(1+\upsilon_1 V_i)(1+\upsilon_{i+1} V_{i+3}) .
\end{equation}
As usual, the RG transformation (4-7) generates new interactions and finally the effective Hamiltonian of the 4-rung block becomes
\begin{eqnarray}
H_r&=& 2A+j\sum_{i=1}^3(S_i S_{i+1}+V_i V_{i+1})+j_p (\frac{1}{2} S_1 V_1+S_2 V_2+S_3 V_3+\frac{1}{2} S_4 V_4) \nonumber \\
&+&j_3 \sum_{i=1}^3 (S_i V_{i+1}+S_{i+1}  V_i)+k \sum_{i=1}^3 S_i S_{i+1} V_i V_{i+1}.
\end{eqnarray}
where parameters of the generated couplings are $ j_3=-\frac{\tilde{J_3}}{T}$ and $ k=-\frac{\tilde{K}}{T}$.

Using formulae (4-7), one can find the renormalized  Hamiltonian in the same shape as the original one with new effective interaction parameters and the RG transformation has the form of five recursion relations for the renormalized parameters $Z_0, J, J_p, J_3$ and $K_4$ as functions of the original ones $A, j, j_p, j_3$ and $k$ (8).
\begin{eqnarray}
Z_0 &=& \frac{1}{8} \log(\lambda_1 \lambda_2 \lambda_3 \lambda_4 \lambda_5^4), \quad  J= \frac{1}{8} \log(\frac{\lambda_2  \lambda_3}{\lambda_1 \lambda_4}), \quad J_p= \frac{1}{4} \log(\frac{\lambda_3  \lambda_4}{\lambda_1 \lambda_2}), \\ \nonumber J_3&=& \frac{1}{8} \log(\frac{\lambda_1  \lambda_3}{\lambda_2 \lambda_4}), \quad  K_4 = \frac{1}{8} \log(\frac{ \lambda_1 \lambda_2 \lambda_3 \lambda_4}{ \lambda_5^4})
\end{eqnarray}
where
\begin{eqnarray}
\lambda_1 &=& e^{2 A -(6 j+2 j_3+3 j_p+k} [2 e^{4(j+j_p)}+4 e^{2( 4j+j_p)}+2 e^{4 (2j+j_3+j_p)}+4e^{2(2j+2 j_3+j_p)} \nonumber \\
&+&3 e^{4(2j+k)}+e^{4(2j_3+k)} ], \nonumber \\
\lambda_2 &=& e^{2 A -(2 j+6j_3+3 j_p+k} [2 e^{4(j_3+j_p)}+2 e^{4( j+2j_3+j_p)}+4 e^{2 (2j+2j_3+j_p)}+4e^{2(4 j_3+j_p)} \nonumber \\
&+& e^{4(2j+k)}+3e^{4(2j_3+k)} ], \nonumber \\
\lambda_3&=& e^{2 A-(2j+2j_3+j_p+k)}[2 e^{4j}+2 e^{4 j_3}+4 e^{2j_p}+4 e^{2(2j+2j_3+j_p)}+3 e^{4(j_p+k)}+e^{4(2j+2j_3+j_p+k)}], \nonumber \\
\lambda_4 &=&e^{2 A-(6j+6j_3+j_p+k)}[2e^{4(2j+j_3)}+2e^{4(j+2j_3)}+4e^{2(2j+2j_3+j_p)}+4e^{2(4j+4j_3+j_p)}+e^{4(j_p+k)} \nonumber \\
&+&3e^{4(2j+2j_3+j_p+k)}], \nonumber \\
\lambda_5&=&e^{2 A-(4j+4j_3+2j_p+3k)}[4e^{2(2j+2j_3+j_p)}+e^{4(2j+k)}+2e^{4(j+j_3+k)}+e^{4(2j_3+k)}+
e^{4(j_p+k)} \nonumber \\
&+&2e^{4(j+j_3+j_p+k)} +e^{4(2j+2j_3+j_p+k)}+e^{2(2j+j_p+2k)}+e^{2(2j_3+j_p+2k)}+e^{2(4j+2j_3+j_p+2k)} \nonumber \\
&+&e^{2(2j+4j_3+j_p+2k)}].
\end{eqnarray}
Evaluating numerically the RG recursion relations (9-10), the free energy per site can be found according to the following formula:
\begin{equation}
f=\sum_{n=1}^{\infty} \frac{Z_0^{(n)}}{3^n} ,
\end{equation}
where $Z_0$ constant (spin independent) term generated in each $n$-th renormalization step,
and then the temperature dependences of the specific heat and nearest-neighbor correlation functions:
\begin{equation}
G_{ss}=<S_i S_{i+1}>, \quad G_{vv}=<V_i V_{i+1}>, \quad G_{sv}=<S_i V_i>, \quad G_{sb} = < S_i B_i>.
\end{equation}

In Fig.2 the dot-marked results obtained for the specific heat from the LRG recursive relations (9-10), are compared with those obtained exactly (solid lines) for three values of the interleg interaction parameter ($j_2=-0.05, -0.1, -0.14$). As seen LRG reproduces exactly the rigorous results as for the standard Ising chain \cite{MN}. To prove this in a general way, one should sum up the infinite series expansion (11) to find the analytical solution \cite{Wei1}, as  M. Nauenberg  \cite{MN} did for the Ising chain.  However, in the latter case the RG transformation reduces to one recursion relation for the interaction parameter, whereas in the case under consideration we have four of them ($j, j_p, j_3$ and $k$) which makes the task much more complicated. But in any case, it seems that LRG is a reasonable approach to study thermodynamic properties of the Ising spin ladders. In the following part we will apply the LRG to the model with next nearest-neighbor interactions which cannot be solved exactly.  Before that, however, in the next section we will discuss the exact solution for an asymmetric Ising spin ladder in an  external magnetic field.

\begin{figure}
\label{Fig_1}
 \epsfxsize=8cm \epsfbox{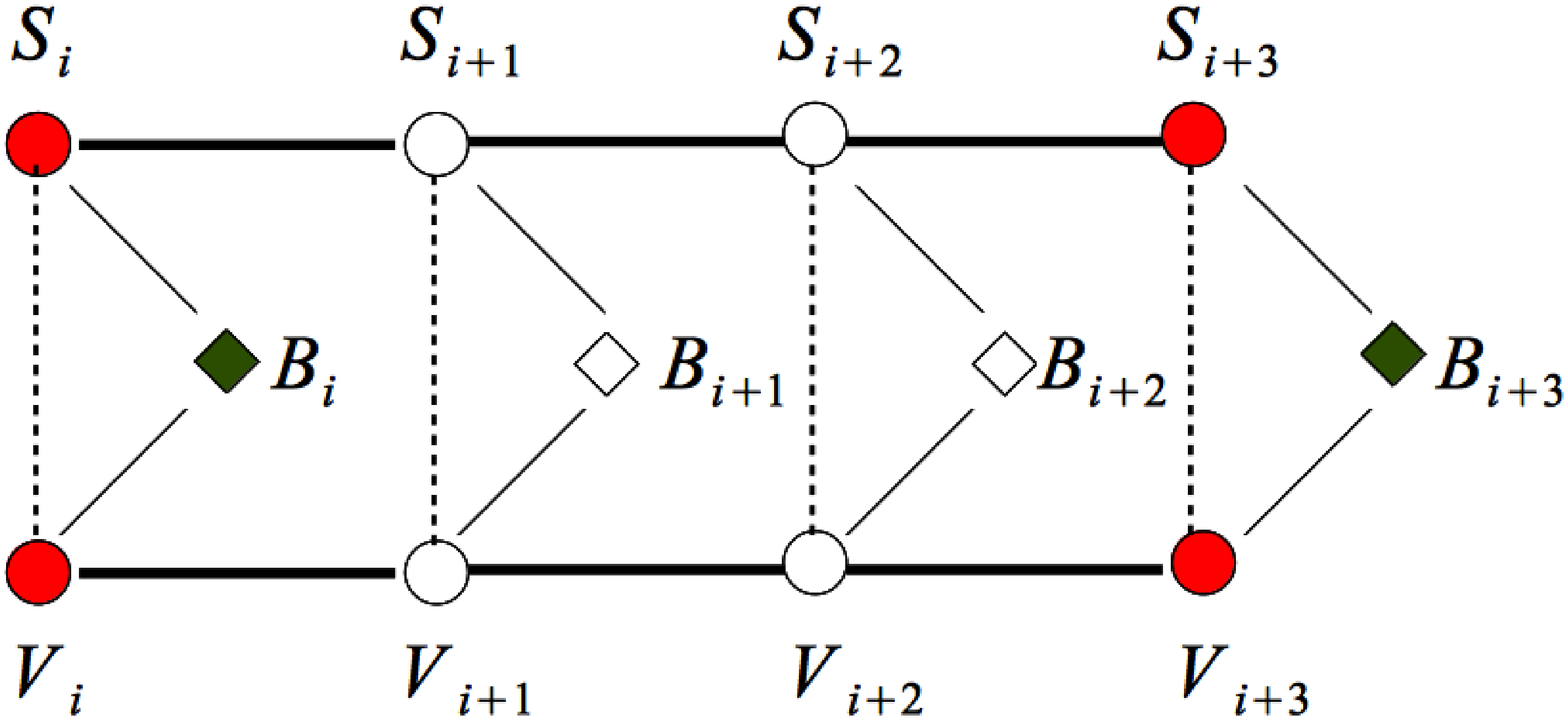}
 \caption{Block used to renormalize two-leg ladder with trimer rungs.}
 \end{figure}

\begin{figure}
\label{Fig_2}
 \epsfxsize=8cm \epsfbox{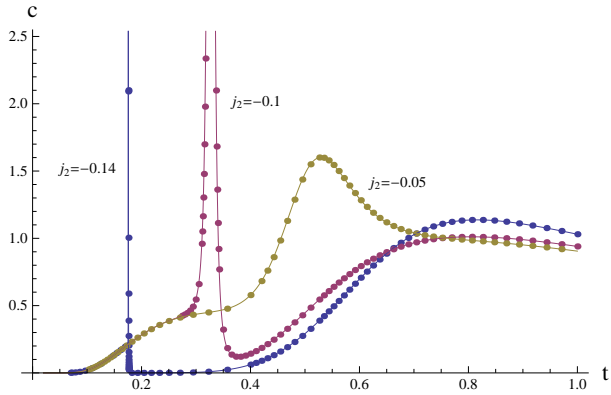}
 \caption{Temperature dependence of the specific heat. Comparison of the results found from the RG transformation (9-10)  (points) with the exact results  (solid lines) \cite{Wei1, Wei2, Hut}.}
 \end{figure}

\section{Asymmetrc Ising spin ladder in a field}

In this section we will consider the two-leg Ising spin ladder with trimer rungs (Fig.1) and different interaction parameters in each leg (asymmetric ladder) in an external magnetic field defined  by the Hamiltonian
\begin{eqnarray}
H_h &=& j_s \sum_i S_i  S_{i+1} + j_v \sum_i V_i V_{i+1} + \frac{1}{2} j_2 \sum_i (S_i V_i + S_{i+1} V_{i+1}) + \frac{1}{2} h_s \sum_i (S_i+S_{i+1})\nonumber \\
&+& \frac{1}{2} h_v \sum_i (V_i+V_{i+1}) + j_{1} \sum_i (S_i + V_i ) B_i + h_b \sum_i B_i.
\end{eqnarray}
Hereafter we name $j_2$ the direct interleg interaction and $j_1$ that couples the middle spin $B_i$ to two leg spins $S_i$ and $V_i$ of the rung,  indirect.
 To calculate the magnetization in the each leg we introduce three fields ($h_s, h_v, h_b$).

Similarly, as in the case without an external field one can perform summation over $B_i$ spins obtaining the effective Hamiltonian in the form
\begin{eqnarray}
H_h& = A_h N+\tilde{H_h}
\end{eqnarray}
and
\begin{eqnarray}
\tilde{H_h}&=& j_s \sum_i S_i  S_{i+1} + j_v \sum_i V_i V_{i+1} + \frac{1}{2} j_{ph} \sum_i (S_i V_i + S_{i+1} V_{i+1}) \nonumber \\
&+& g_s \sum_i (S_i+S_{i+1})+g_v \sum_i (V_i+V_{i+1}).
\end{eqnarray}
where
\begin{eqnarray}
j_{ph} &=& j_2+ T B_h, \quad g_s = \frac{1}{2}  h_s +T G, \quad  g_v = \frac{1}{2}  h_v +T G, \nonumber \\
A_h&=&\frac{1}{4} \{ \log[2\cosh (\frac{h_b}{T}-\frac{2j_1}{T})] +\log[2\cosh (\frac{h_b}{T}+\frac{2j_1}{T})]+2\log[2\cosh (\frac{h_b}{T})] \},\nonumber \\
B_h&=&\frac{1}{4} \{ \log[2\cosh (\frac{h_b}{T}-\frac{2j_1}{T})] +\log[2\cosh (\frac{h_b}{T}+\frac{2j_1}{T})]-2\log[2\cosh (\frac{h_b}{T})] \}, \nonumber \\
G=&=&\frac{1}{4} \{ \log[2\cosh (\frac{h_b}{T}+\frac{2j_1}{T})]-2\log[2\cosh (\frac{h_b}{T})] - \log[2\cosh (\frac{h_b}{T}-\frac{2j_1}{T})]  \}.
\end{eqnarray}

Introducing 4 $ \times$ 4 transfer matrix {\bf P}
\\

$\left ( \begin{array}{c} P_0 \quad P_1 \quad P_2 \quad P_3 \\ 
 P_1 \quad P_4 \quad P_5 \quad P_6 \\
 P_2 \quad P_5 \quad P_7 \quad P_8 \\
 P_3 \quad P_6 \quad P_8 \quad P_9 \end{array}  \right ),$
 \\
 where
 \begin{eqnarray}
P_0 &=& e^{(2g_s+2g_v+j_s+j_{ph}+j_v)/T}, \quad P_1=e^{(2 g_v-j_s+j_v)/T}, \quad P_2=e^{(2 g_s+j_s-j_v)/T}, \quad P_3=e^{(-j_s+j_{ph}-j_v)/T}, \nonumber \\
P_4&=&e^{(2g_s-2 g_v+j_s-j_{ph}+j_v)/T}, \quad P_5=e^{(-j_s-j_{ph}-j_v)/T},  \quad P_6=e^{(-2 g_s+j_s-j_v)/T}, \nonumber \\
P_7&=&e^{(-2 g_s+2 g_v+j_s-j_{ph}+j_v)/T}, \quad P_8=e^{(-2 g_v-j_s+j_v)/T}, P_9=e^{-2 g_s-2 g_v+j_s+j_{ph}+j_v)/T},
\end{eqnarray}
 one can easily find by using Mathematica the analytical but rather complicated form for the largest eigenvalue of {\bf P},  $L_{max}$.
 This gives for the free energy per rung in the thermodynamic limit
 \begin{equation}
 f = -T A_h-T \log L_{max}
 \end{equation}
 and then specific heat, leg magnetization and correlation functions: $G_{sv} = <S_i V_i>$ and $G_{sb}= <S_i B_i>$.
 
 \subsection{$h=0$}
 
Let us start with the standard two-leg ladder with trimer rungs at zero field.  In Fig.3 two possible shapes of the specific heat curve are presented.
In both cases we assume $j_v = j_s =1$. For $j_1=0.07$ the three peak structure of the specific heat is observed (Fig.3 left column) for both the non-frustrated  ($j_2 =0$) and the frustrated  ($j_2 < 0$) model.
From high temperature, the first broad maximum refers to a large change in the mutual setting of spins in several legs, the second to 
linking of the spins between legs ($G_{sv}$) and finally a low temperature peak related to the
linking $S_i$ and $B_i$  spins ($G_{sb}$ ). 
It is clearly seen in Fig.4 where the specific heat extremes are related to the inflection points of the several correlation functions marked by their second derivatives (thin lines).
For negative direct interleg couplings  $j_2$ the maxima become more pronounced and the effective interleg coupling $j_p$ changes sign at a certain temperature $ T_p $. As was shown \cite{Hut} vanishing of $j_p$ is a necessary condition for the existence of the phase crossover under discussion.  As seen with decreasing of $j_2$ the temperature $ T_p $ decreases and for some value of $j_2=j_2^*$
coincides with the temperature of the function $G_{sv} $ inflection point, this is the phase crossover temperature $T_p^*$. Then, for $j_2 < j_2^*$ both temperatures decreases together up to $j_2 = -j_1$. For $T < T_p^*$ the system is non-frustrated ($G_{sv} > 0$) and at $T>T_p^*$ becomes frustrated. So, at $T=T_p^*$ the system  changes from non-frustrated to frustrated. Thus, if the temperature at which the system changes the status from non-frustrated to frustrated coincides with the temperature at which the spins from the both legs become 
unbound (indicated by the second maximum of the specific heat), 
then the mentioned peculiar behavior of the thermodynamic quantities are observed. For larger value of $j_1$ ($j_1=0.2$) the two high temperature maxima of the specific heat overlap but the anomaly scenario is the same.

\begin{figure}
\label{Fig_3}
 \epsfxsize=15cm \epsfbox{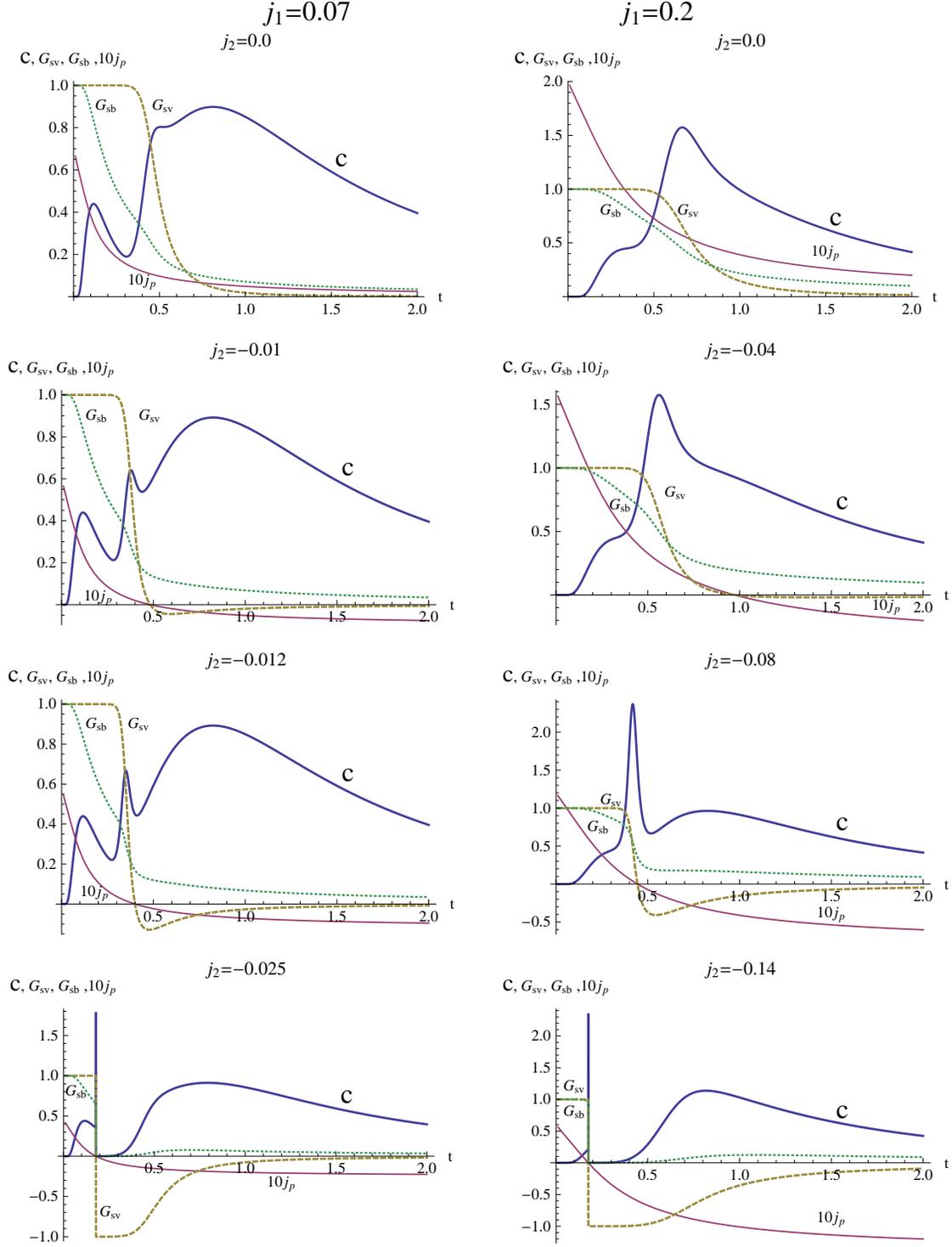}
 \caption{Temperature dependences of the specific heat (solid bold lines), correlation $G_{sv}$ (dashed lines), $G_{sb}$ (dotted lines) and effective interleg interactions $j_p$ (multiplied by $10$) for symmetric ladder $j_{v}  =  j_{s}$, two values of the indirect  interaction $j_1 = 0.07$ and $0.2$ and several values of the direct interleg interaction $j_2$ at $h=0$.}
 \end{figure}

\begin{figure}
\label{Fig_4}
 \epsfxsize=10cm \epsfbox{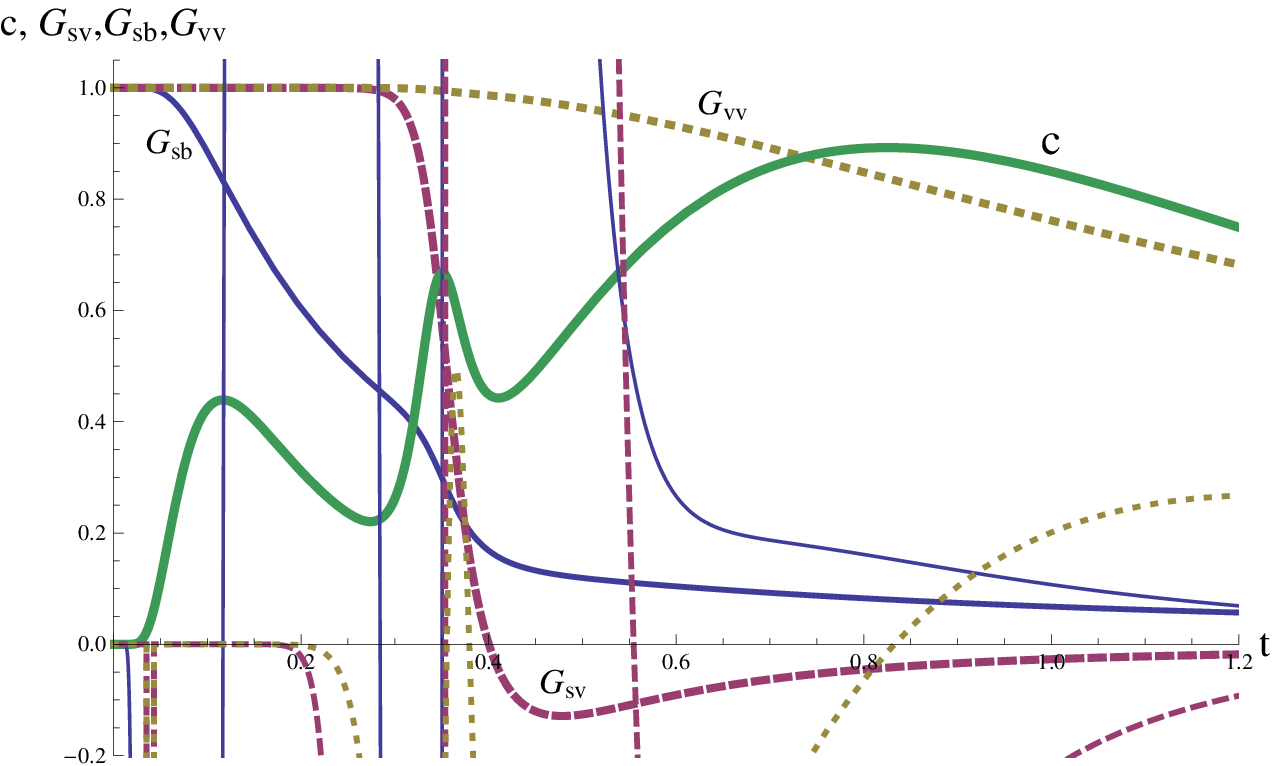}
 \caption{Temperature dependences of the specific heat (solid bold line) and correlation functions $G_{sb}$ (solid line), $G_{sv}$ (dashed), $G_{vv}$ (dotted) and their second derivatives: thin, thin-dashed and thin-dotted, respectively, for the ladder with $j_1=0.07$ and $j_2=-0.012$.}
 \end{figure}

For an asymmetric ladder with different ferromagnetic interactions in both legs, for example $j_v < j_s$ the anomaly survives although the
crossover point is shifted towards the lower temperature Fig.5 ($j_v=0.1$). For the antiferromagnetic interaction in one of the legs, the effect is
destroyed Fig.5  ($j_v=-0.1, j_v=-1$), though there is still a temperature at which the effective coupling $j_p$, independent of  the intraleg interactions, vanishes.

\begin{figure}
\label{Fig_5}
 \epsfxsize=17cm \epsfbox{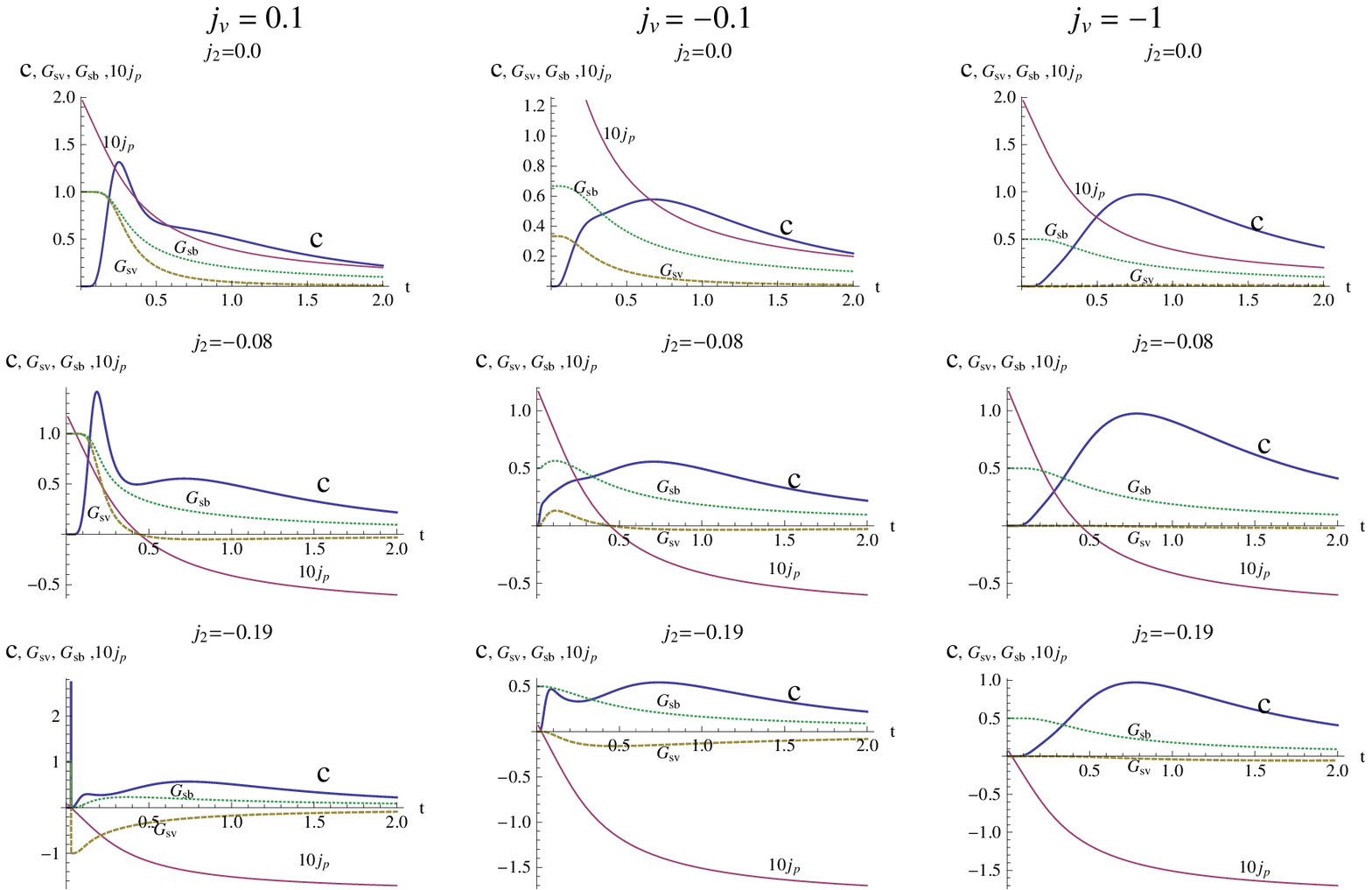}
 \caption{The same as in Fig.3 for asymmetric ladder $j_{v}  \neq  j_{s}$ with indirect interleg coupling $j_1=0.07$ and $j_v = 0.1 j_s$ left column; $j_v = -0.1  j_s$ middle column and  $j_v =- j_s$ (right column) at $ h=0$ for several values of the direct interleg coupling $j_2$.}
 \end{figure}
 
 \subsection{$h \neq 0$}
 
Since there is no real phase transition, the external field does not destroy the effect but only melts it, as was shown for Ising diamond and tetrahedral chains \cite{Str2}. In Fig.6 the temperature dependences of the several leg magnetization for three values of the field are presented.
The upper row refers to the symmetric case $j_v=j_s$ and the bottom row to the asymmetric one with $j_v=0.6 j_s$. As seen for the symmetric case the clear crossover effect indicated by the very abrupt jumps of both leg magnetizations ($m_s, m_v$) from "zero" to the "saturation" ($m_i=1$)  for a sufficiently small field occurs. For a higher field the effect is smeared then disappears. For the asymmetric case and very small field only the magnetization of the leg with stronger intraleg  interaction $m_s$  ($j_s > j_v$) undergoes the jump, but from the value close to $-1$ to $1$. For the higher field $h=0.1$  again the sharp crossover decays. Although the maximum value of the field  for which $j_p$ changes sign at a finite temperature is $h=h_{max}=0.12$. (for used set of the interaction parameters $j_s=1, j_v=0.6, j_1=0.2$ and $j_2=-0.14$).

\begin{figure}
\label{Fig_6}
 \epsfxsize=17cm \epsfbox{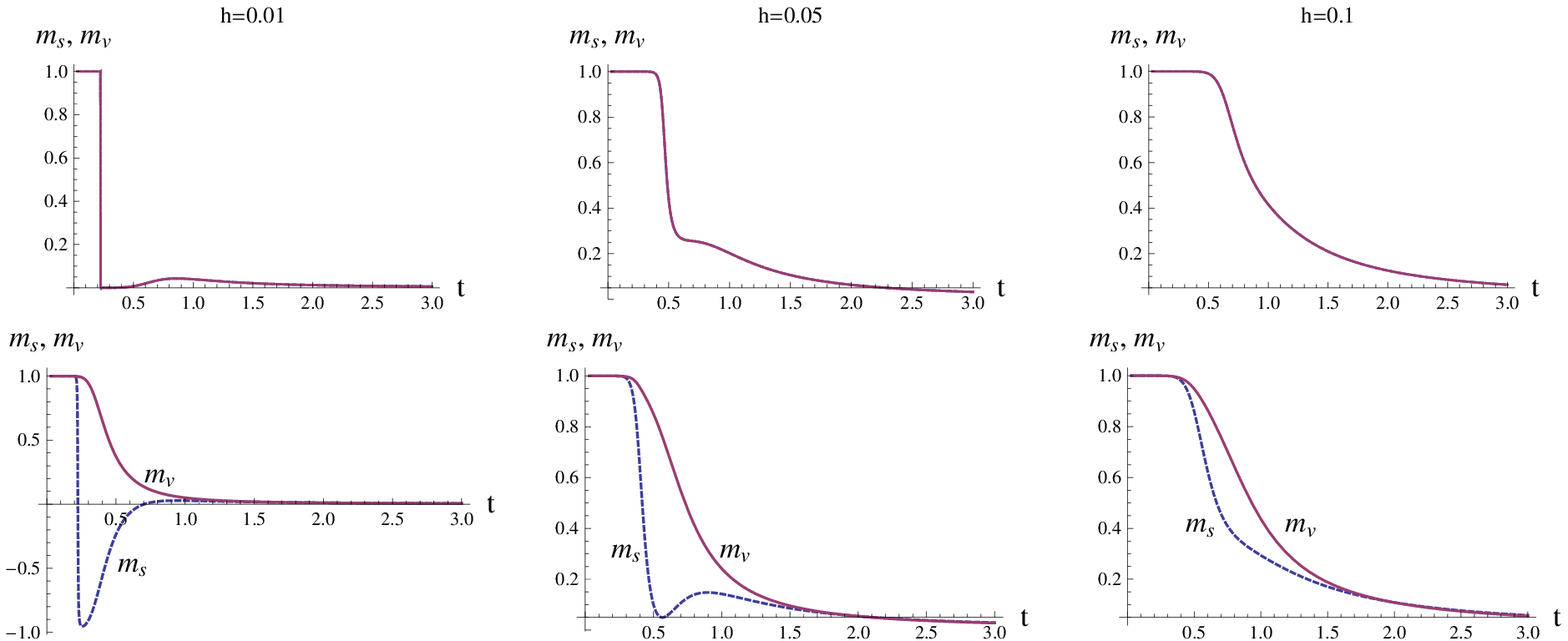}
 \caption{Temperature dependences of the leg magnetizations $m_s$ (dashed lines) and $m_v$ (solid lines) of the  ladder with $j_s=1, j_1 =0.2$ and $j_2 = -0.14$ for several values of the field $h=0.01, 0.05$ and $0.1$ with  $j_v = j_s $ (upper row) $j_v=0.6 j_s$ (bottom row).}
 \end{figure} 
 
\begin{figure}
\label{Fig_7}
 \epsfxsize=12cm \epsfbox{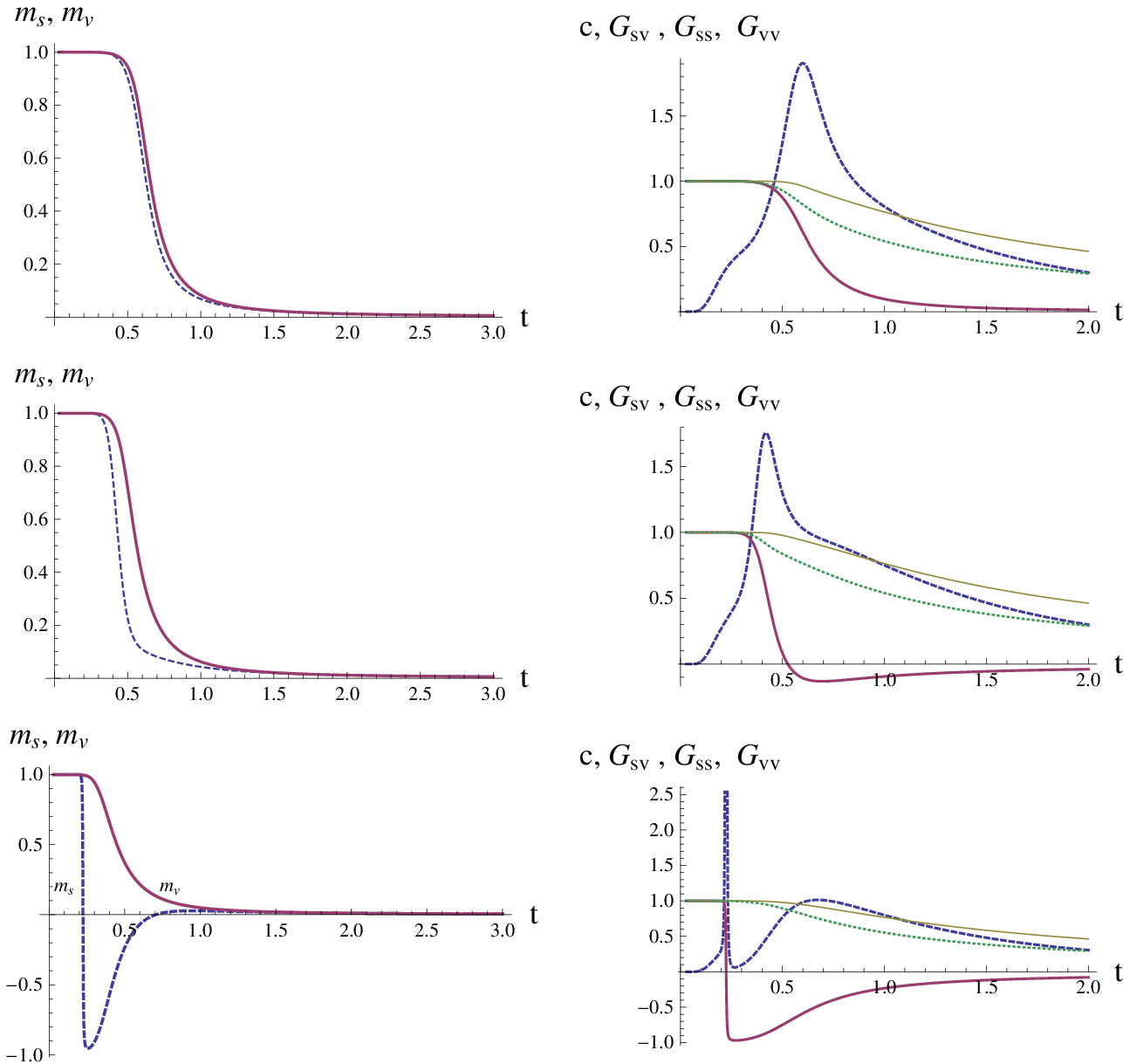}
 \caption{Left column: Temperature dependences of the leg magnetizations $m_s$ (dased lines) and $m_v$ (solid lines)  and right column: Specific heat (dashed lines) and correlations  $G_{sv}$ (solid lines), $G_{ss}$ (thin) and $G_{vv}$ (dotted)  for the asymmetric  ladder with $j_s=1, j_v=0.6, j_1 =0.2$ at $h_s=h_v=h=0.01$ and $j_2 = 0.0, -0.08, -0.14$ from top to the bottom.}
 \end{figure}
 
In Fig.7 The temperature dependences of the leg magnetization (left column) and the specific heat, intraleg correlations $G_{ss}, G_{vv}$, and interleg correlation function $G_{sv}$ for $h=0.01$ and several values of the direct interleg interaction are shown.

\section{Ladder with NN and NNN interactions.}

In order to study a system with next-nearest-neighbor interactions one has to use the LRG projector (5) which survives three spins in a leg (full circles and full squares in Fig. 8) instead of two as was the case with only nearest-neighbor interactions (6) and now the linear RG  projector is given by
\begin{equation}
 p_i(\sigma, \upsilon; S,V) =\prod_{n=0}^{2} (1+\sigma_{i+n} S_{i+2n})(1+\upsilon_{i+n} V_{i+2n}).
\end{equation}
In this case the LRG transformation generates 10 new interactions and the final Hamiltonian includes in addition to the original couplings NN and NNN ($j_s, j_v, j_{2s}, j_{2v}$ and $j_p$) all possible two-, four- and six-spin interactions acceptable by the symmetry of the renormalized block (see appendix).

In our previous paper \cite{JS1} to renormalize the two-spin model with both NN and NNN interactions we have used seven site cluster and confined ourselves only to the bilinear terms in the Hamiltonian. Now we use five-rung cluster (Fig.8) and keep all terms of the renormalized Hamiltonian and the LRG transformation has a form of 16 recursion relations for the parameters $Z_0, J_s, J_v, J_{2s}, J_{2v}, J_p,  J_3, J_4, K_1, K_2, K_3, K_5, K_7, K_8, K_9$ and $R_6$ as functions of the original ones (20). In the standard way (see appendix in Ref.  \cite{JS1})  one can find the analytical relations between renormalized and original parameters although they have rather complex form. Evaluating numerically the LRG recursion relations one can find the free energy according to the formula (11) but with the scaling factor $2$ instead of $3$. So, the energy per site for the cluster in Fig.8 with every second site (empty circles and squares) killed in the LRG step is given by
\begin{equation}
f=\sum_{n=1}^{\infty} \frac{Z_0^{(n)}}{2^n} .
\end{equation}
As shown in section II for a ladder with only NN interactions ($j_{2i}=0$) the LRG leads to the known exact results. Unfortunately, the models with NN and NNN interactions are not exactly solvable and there are not results to be compared.
 Using the formula (20) one can calculate  the specific heat and correlation functions $G_{vv}, G_{sv}$  and $G_{sb}$ (12).

\begin{figure}
\label{Fig_8}
 \epsfxsize=8cm \epsfbox{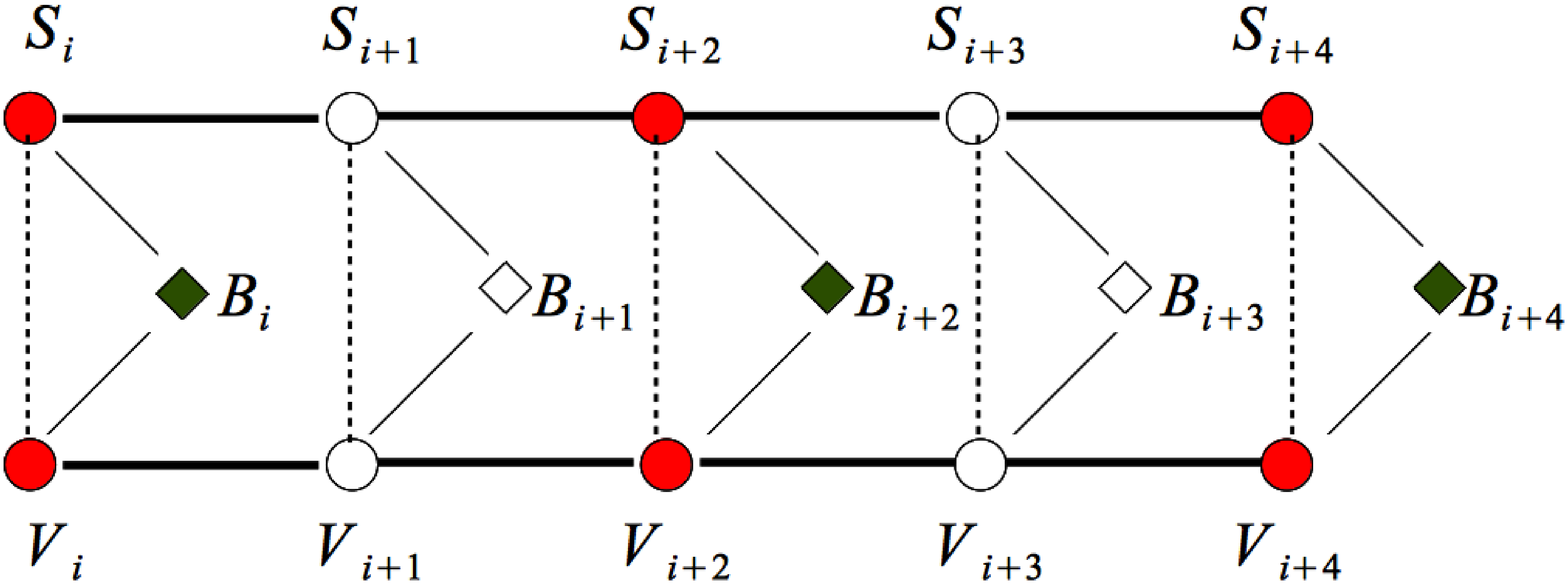}
 \caption{The cluster used to get renormalized Hamiltonian of the Ising  spin ladder with trimer rungs and NNN interactions. The open circles and open squares are eliminated in each step of the LRG transformation.}
 \end{figure}

\begin{figure}
\label{Fig_9}
 \epsfxsize=17cm \epsfbox{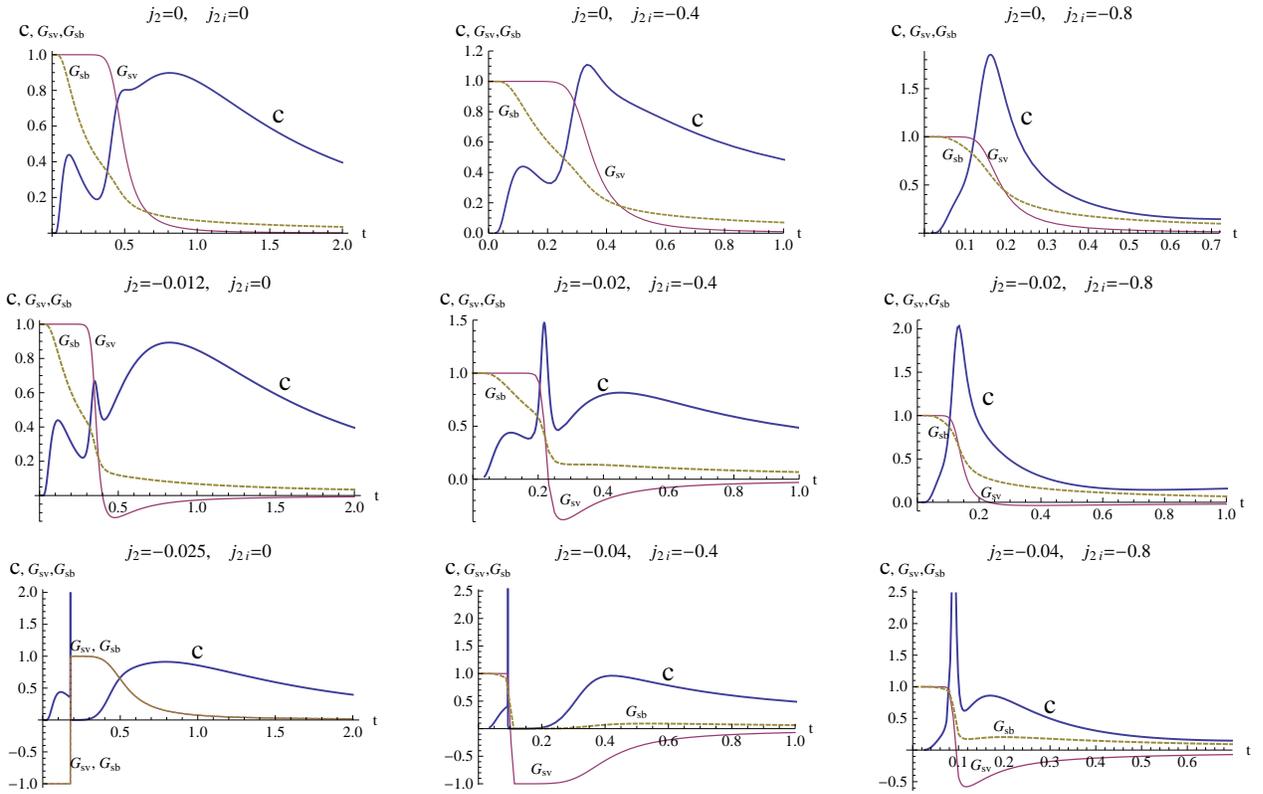}
 \caption{Temperature dependences of the specific heat and correlation functions for the symmetric ladders with indirect coupling $j_1=0.07$ with only NN interactions (left column), with both NN and NNN interactions $j_{2s}=j_{2v}=-0.4$ (middle column)  and  $j_{2s}=j_{2v}=-0.8$ (right column) for three values of direct coupling $j_2$.}
 \end{figure}

Fig.9 shows the temperature dependences of these quantities for the symmetric spin ladder with indirect interaction $j_1=0.07$,  NNN interactions $j_{2s}=j_{2v}=j_{2i}=0, -0.4,-0.8$ and several values of the direct interleg coupling $j_2$. As can be seen, NNN interactions significantly change the shape of the heat curve in particular for large values of $\mid j_2 \mid $ ($j_2=-0.8$), where only one maximum of the specific heat is visible instead of three for $j_2=0$.
However, for a sufficiently high absolute value of the direct interleg coupling $j_2$ the phase crossover (very sharp maximum of the specific heat and jump of the correlation functions) still occurs. 
In Fig.10 two classes of models, symmetric ($j_s=j_v=1,  
j_{2s}=j_{2v}=-0.4$) and asymmetric ($j_s=1, j_v=0.6, j_{2s}=0, j_{2v}=-0.4$) are compared. Clear differences can be seen between the heat and the correlation functions curves for both systems, but in both cases there is a clear link between the specific heat maximum and the inflection point of the correlation functions $G_{sv}$ and $G_{sb}$. For the asymmetric ladder also the intraleg correlation $G_{vv}$ exhibits an inflaxtion point at the same temperature, in contrast to $G_{ss}$.
The temperature at which both correlation functions $G_{sv}$ and $G_{sb}$ exhibit the inflection point  (the largest maximum of specific heat) is denoted by $c_{max}$, and the temperature at which the effective interleg coupling $j_p$ vanishes is denoted by $j_p^0$. In the bottom panel of the Fig.10 the dependences of  $c_{max}$ and $j_p^0$ on the direct interleg interaction are presented. As already mentioned in section III at some value of $j_2$ both curves coincide, indicating the phase crossover.

\begin{figure}
\label{Fig_10}
 \epsfxsize=13cm \epsfbox{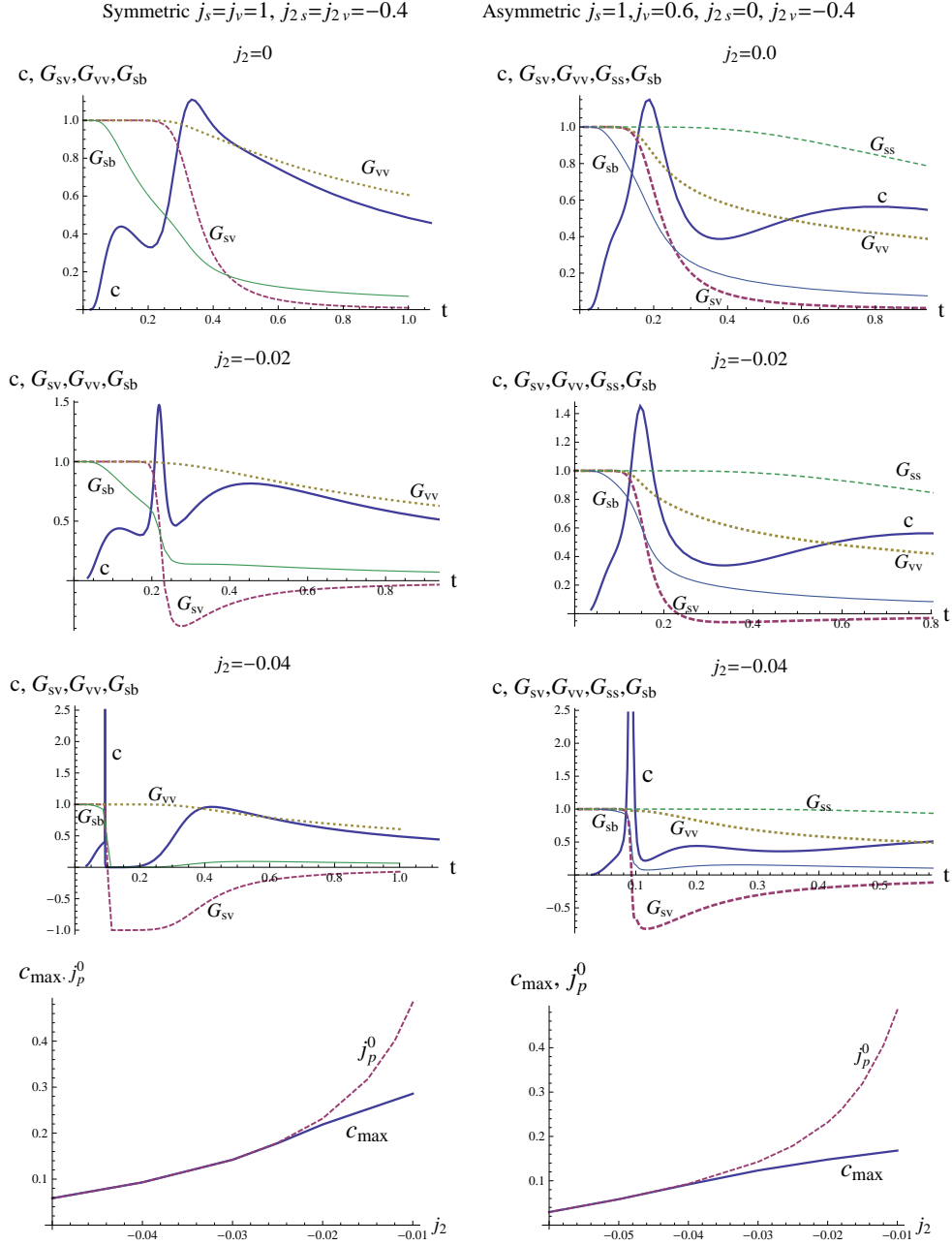}
 \caption{Comparison of the thermodynamic quantities of the symmetric and asymmetric ladders with NN and NNN interactions.The temperature dependences of the specific heat (bold lines) and correlation functions $G_{sv}$ (dashed), $G_{sb}$ (thin) and $G_{ii}$ (dotted) for three values of the direct interleg coupling $j_2 = 0, -0.02, -0.04$. The bottom panel shows the temperature point of change of the effective interleg interaction parameter $j_p^0$ (dotted line) and location of specific heat maximum related with the $G_{sv}$ inflection point $c_{max}$ (solid line).}
 \end{figure}

\section{ACLS model with frustration}

A Hamiltonian formulation of social dynamics has a long history. Several Hamiltonian terms are interpreted as various interpersonal  interactions and  an environment (family, friends, mass media) influence. Such, most often Ising like, Hamiltonians were used to determine the eventual equilibrium states of agent systems. Depending on the process under investigation, the Hamiltonian was called the global dissatisfaction function \cite{Galam} or the disagreement function \cite{KW1, KW2} etc, and an additional assumption was made to justify minimization of these functions, e.g., the minimum dissatisfaction principle \cite{Galam}. 
Applying the principles of statistical physics to such a Hamiltonian,  mostly in the  molecular field approximation, all features of the Ising model, including a phase transition, are revealed. It is clear that such a procedure is not fully justified for the dynamical agent models,  but it should define the stable states to which  the dynamical model converges. To be sure in a realistic model such a state is never achieved but its knowledge can indicate the direction of a system evolution.

As known, a true phase transition can only appear in the thermodynamic limit.  Nevertheless,  several sociophysical models  composed by a finite number of agents, show behavior that resemble a phase transition. This means that a change in behavior can be observed on both sides of a certain point $p_c$. 
However, Toral and Tessone \cite{Toral} have shown, that  in some models of relevance to social systems, " the value of $p_c$ depends on the system size in such a way that $p_c$ tends to zero when the system size N increases to infinity. Therefore, in a strict thermodynamic sense, the transition does no exists since it does not survive the thermodynamic limit."
Nonetheless, the character  of the phase transitions  (continuous or discontinuous) and even the critical singularities (critical indices) are subject of study in many agent models.  Especially, that an abrupt change in society status (opinion), which can be interpreted as a \emph {pseudo-transition} in some real population is also observed \cite{Toral}. Perhaps, in some cases the sudden change of the society views can be described not by controversial in a way an order-disorder transition but by the crossover frustrated - non-frustrated state discussed in section III.A. 

In our previous paper we have considered the Hamiltonian formulation of the dynamical model simulating the process of opinion spreading in a society, where people adopt one of the four social attitudes \cite{JS1}, the ACLS model \cite{KW}. In the present notation such a Hamiltonian reads

\begin{equation}
H_{ACLS} = \sum_i ( j_s S_i S_{i+1} +j_v V_i V_{i+1} ) + j_{2v} \sum_i V_i V_{i+2} + j_2 \sum_i S_i V_i,
\end{equation}
where $j_s$ (interaction in $S$-spin leg) refers to convincing force in the personal area [standard Ising interaction between nearest neighbors (NN)] and $j_v, j_{2v}$  (interactions in $V$-spin leg) refers to convincing forces in the economic area. Direct interleg interaction $j_2$ describes a mutual influence of the both areas. It has been shown \cite{JS1} among others that interleg interaction $j_2$ can trigger or remove a specific heat maximum which can indicate a kind of local (short range) reordering. The process is clearly smooth and does not resemble any phase transition. However, if the Hamiltonian  (21) will be completed with an indirect interleg interaction term
\begin{equation}
j_1 \sum_i B_i (S_i + V_i),
\end{equation} 
then the method and results of the previous section can be used to explain  behavior that brings to mind the phase transition.
Where does such a term come from? There are some evidences that the political views are shaped by personality traits. Admittedly, the question of whether and to what extent these traits can be changed by life experiences is a moot point,  however, they cannot be changing ad hoc by interacting with other agents, therefore there is no interaction between $B_i$ spins.  

Let us consider three cases related to the model of political opinion formation \cite{KW, JS1}. In all of them we take $j_s =1, j_1=0.2, j_2=-0.14$ and:

(i) $j_v = 1, j_{2v} = 0$ (related to the dynamical agent model  where in both areas - personal and economic,  the information flows inward from the neighborhood, like in the most opinion dynamic models \cite{Hol}), 

(ii)  $j_v = 1, j_{2v} = 1$ (\emph{if you do not know what to do, just do nothing}) \cite{KW0} ,

(iii) $j_v = 0, j_{2v} = 1$ (\emph{USDF - united we stand divided we fall}) \cite{KW0}. 

In the two latter cases  the information flows inward from the neighborhood in the personal area ($S$ subsystem) whereas in the economic area  ($V$ subsystem) the information flows outward to the neighborhood.
In Figs.11 and 12 the temperature dependences of the specific heat and correlation functions for these models are presented. As has been postulated in our previous paper \cite{JS1} in politics a response function which is a counterpart of the specific heat should describe the sensitivity of the political system on dissatisfaction of the agents. As seen (Fig.11) in all cases this sensitivity at certain point changes in an extreme way which resembles a phase transition, simultaneously the correlation function $G_{sv}$ exhibits a jump. For the two first cases (i) and (ii) the shapes of the curves are very similar whereas  in the third case where $j_v=0$ the shapes  are clearly different.
There is an additional maximum at low "temperature" related to an inflexion point of  $G_{sv}$  which is associated with
degeneration in the $V$-spin subsystem. Fig.12 shows the temperature dependences of the  intraleg correlation function
$G_{vv}$, interleg $G_{sv}$  and leg - rung spin correlation $G_{sb}$.  
Again for the two first models the shapes of the curves are almost identical. The process of the spin ordering  in the leg is smooth ($G_{vv}$) 
and between legs is sharp ($G_{sv}$). 
Nearby the crossover point from the high temperature side $G_{sv} \approx -1$ and $G_{sb}$ vanishes due to the frustration and below the crossover point both functions abruptly reach the value $\approx 1$. 
This is not the case for the third model, where $G_{vv}$ vanishes at the crossover point and interleg spin arranging
takes place in two steps. The first is sharp,  $G_{sv}$ changes from $\approx -0.5$ to $\approx 0.5$ and the next step is smooth with an inflection point related to the maximum of the specific heat (Fig.11).  These two scenarios should also lead to different dynamics of the relevant agent model.

The internal conflict between different areas of attitudes, personal and economic modulated by the environment (other agents, mass media or social organization) embodied by the direct interaction $j_2 S_i V_i$ can lead to a smooth change of the dissatisfaction level \cite{JS1} and consequently to instability of the political system. However, if one takes into account that personality traits independent of the current environment  [represented by $j_1 B_i (S_i + V_i)$] can also influence the political views that cause  frustration, then the change can be very sharp (revolutionary). To analyze this suggestion, one should of course postulate a relevant dynamic rules as it was done for ACLS model \cite{KW1}.

\begin{figure}
\label{Fig_11} 
 \epsfxsize=15cm \epsfbox{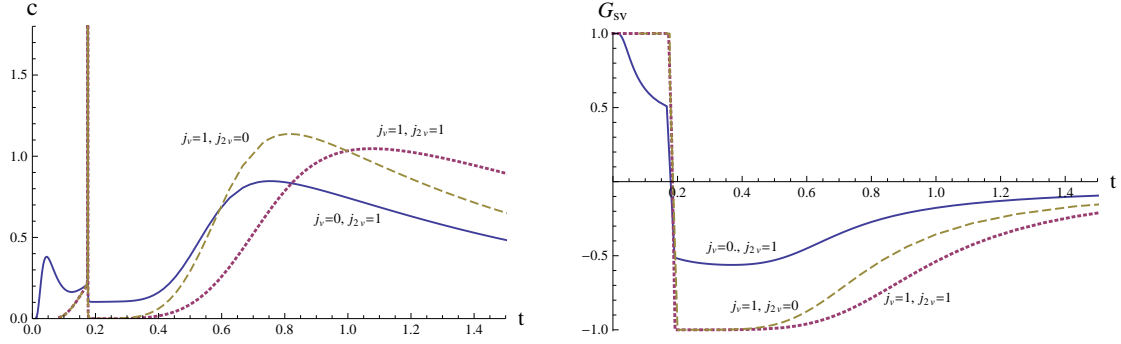}
 \caption{Temperature dependences of the specific heat and correlation function $G_{sv}$  for three models with $j_s=1$ and (i) $j_v=1, j_{2v}=0$ (dashed lines), (ii) $j_v=1, j_{2v}=1$ (dotted) and (iii) $j_v=0, j_{2v}=1$ (solid).}
 \end{figure}
 
\begin{figure} 
\label{Fig_12}
 \epsfxsize=17cm \epsfbox{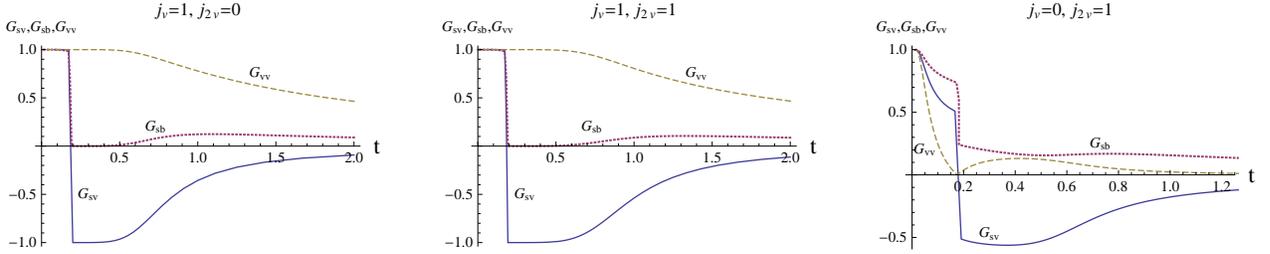}
 \caption{Temperature dependences of the correlation functions  $G_{sv}$ (solid lines),  $G_{vv}$ (dashed) and $G_{sb}$ (dotted)
 for three models   (i) $j_{v} =1, j_{2v}=0$, (ii)  $j_{v} =1, j_{2v}=1$ and  (iii) $j_{v} =0, j_{2v}=1$.}
 \end{figure}

\section{Conclusion}

We have studied the one dimensional  Ising spin model made of two chains (legs) allowing different spin-spin  nearest- and next-nearest-neighbor interactions in each chain. The chains are coupled by the direct interaction ($j_2$) and  indirectly ($j_1$) via an additional spin between them. Such a spin arrangement - the ladder with trimmer rungs - in the standard version with the identical chains (symmetric ladder), has been studied in many papers \cite{ Wei1, Wei2, Hut, Tsv, Krok}. The reason is that such a system exhibits  a very interesting effect - a frustration driven very sharp phase crossover at finite temperature which resembles a phase transition, impossible in one dimension.

First, we have shown that linear renormalization group (LRG)  for the standard two-leg Ising spin ladder with trimer rungs  exactly reproduces the known rigorous results found by using the transfer matrix (TM) method \cite{Wei1, Wei2, Hut}. 
In the present paper, the TM method has been used to study a  ladder with NN different interactions in each leg (asymmetric ladder) at an external magnetic field. At zero field, the specific heat curve of the symmetric ladder exhibits two or three maxima associated with the inflection points of the correlation functions: interleg $G_{sv}$, intraleg $G_{vv}=G_{ss}$ and leg-rung $G_{sb} = G_{vb}$ (Fig.3). 
As was shown  \cite{Hut} the necessary  condition for the existence of the sharp phase crossover  is vanishing of the effective interleg interaction $j_p$ (3) at finite temperature as a result of direct and indirect interactions  competition of the spins from the different legs. However, it is not a sufficient condition, additionally, the vanishing $ j_p$ must coincide with the inflection point of the interleg correlation function $G_{sv}$ (maximum of the specific heat). In other words, the inflection point of the interleg correlation function has to coincide with its change of the sign. This means that above the crossover point, $G_{sv} < 0$ and the system is to some extent frustrated and below, $G_{sv} > 0$ and the system is non-frustrated. For the asymetric ladder, the shape of the specific heat curve is different but the sharp crossover survives provided that the intraleg couplings in both legs are positive (ferromagnetic). If the intraleg interaction in one of the leg is negative (antiferromagnetic), the phenomenon does not appear (Fig.5), although the temperature at which the effective interleg coupling disappears still exists. It is clear because this coupling is independent of the intraleg interactions (16). 

The sharp effect also exists in the presence of a sufficiently small external magnetic field for both symmetric and asymmetric ladder. By introducing three fields conjugated to the respective group of spins ($S_i, V_i, B_i$) we have found the magnetization of each leg. At the crossover point only the magnetization of the leg with larger intraleg coupling $m_s$ (for $j_s > j_v$) exhibits a sharp jump  whereas the other one $m_v$ changes smoothly (Fig.7). Also the intraleg correlation functions $G_{ii}$ are smooth at the crossover point and only interleg one $G_{sv}$ shows a jump. Unlike the ladder with only NN interaction, the model with both NN and NNN interactions cannot be solved exactly. Therefore, we have used the LRG method to consider its thermodynamic properties.  We have studied the models with positive (ferromagnetic) NN and negative (antiferromagnetic) NNN interactions.  NNN interactions change the shapes of the specific heat and correlation functions curves as functions of temperature. However, in both symmetric and asymmetric case the sharp phase crossover survives. It has been shown that in all cases this crossover takes place when the point  at which the effective interleg coupling $j_p$ vanishes ($j_p^0$) meets the point at which the interleg correlation function exhibits an inflection point ($c_{max}$) accompanied by the specific heat maximum (Fig.10).

It should be emphasized that in all cases considered, as in previously published papers \cite{Wei1, Wei2, Hut, Rojas} for other models, the observed very sharp crossover accompanied by a gigantic increase of the specific heat is not a true phase transition. That is, the free energy is an analytic function in the thermodynamic limit, although the specific heat exhibits a peak of extreme height.

Finally, we have shown that in some cases when studying political processes in a democratic system it is not necessary to refer 
to the concept of a phase transition to describe  an abrupt change in the views of a society. Such a change can 
be described by the phase crossover triggered by frustration caused by  internal conflicts between different areas of  attitudes, 
personal and economic of individual members of society,  if in addition their personality traits are taken into account.
Of course, in order to obtain practical and credible conclusions, it is necessary, first of all, to formulate rules defining opinion dynamics corresponding to the assumptions of the proposed model (21, 22) as was done for the ACLS model \cite{KW}.

\section{Acknowledgement }
I am grateful to Oleg Derzhko for inspiring talks.

\section{Appendix}

For the five-rung cluster (Fig.8) relating to the projector (19) the complete Hamiltonian takes the form
\begin{eqnarray}
H_r &=& 2A +  \sum_{i=1}^4( j_s S_i S_{i+1}+ j_v V_i V_{i+1}) + \sum_{i=1}^3( j_{2s} S_i S_{i+2}+ j_{2v} V_i V_{i+2})  \nonumber \\
&+&  j_p (\frac{1}{2} S_1 V_1+S_2 V_2+S_3 V_3 + S_4 V_4+\frac{1}{2} S_5 V_5)
+j_3 \sum_{i=1}^4 (S_i V_{i+1}+S_{i+1}  V_i) \nonumber \\
&+& j_4 \sum_{i=1}^3 (S_i  V_{i+2}+ S_{i+2} V_i) + k_1  \sum_{i=1}^4 S_i S_{i+1}  V_i V_{i+1} + k_2  \sum_{i=1}^3 S_i S_{i+2}  V_i V_{i+2}  \nonumber \\
&+& k_3 \sum_{i=1}^3 S_i S_{i+1}  S_{i+2} (V_i + V_{i+1} +V_{i+2}) +  k_5 \sum_{i=1}^3 V_i V_{i+1}  V_{i+2} (S_i + S_{i+1} +S_{i+2})  \nonumber \\
&+& k_7 \sum_{i=1}^3 ( S_i S_{i+1} V_{i+1} V_{i+2} + S_{i+1} S_{i+2} V_i V_{i+1}) + k_8  \sum_{i=1}^3 S_i S_{i+1} V_i V_{i+2}   \nonumber \\
&+&  k_9  \sum_{i=1}^3 S_i S_{i+2} V_i V_{i+1} + r_6  \sum_{i=1}^3 S_i S_{i+1} S_{i+2} V_i V_{i+1} V_{i+2}.
\end{eqnarray}
As mentioned in our previous paper \cite{JS1} the LRG transformation with the projector (19) drives the inequivalence of the renormalized spins at sites ($i, i+2$) and $i+1$. So, using the transformation (4,5) with the projector (19) one finds the renormalized Hamiltonian that in the shape differs slightly from the original one
\begin{eqnarray}
H'_r &=& Z_0 +  \sum_{i=1}^2( J_s \sigma_i \sigma_{i+1}+ J_v \upsilon_i \upsilon_{i+1}) + J_{2s} \sigma_i \sigma_{i+2}+ J_{2v} \upsilon_i \upsilon_{i+2}  \nonumber \\
&+&  J_p (\frac{1}{2} \sigma_1 \upsilon_1+\sigma_3 \upsilon_3 )+J_{pc} \sigma_2 \upsilon_2 + 
J_{3s}  (\sigma_i +\sigma_{i+2})  \upsilon_{i+1} + J_{3v}  (\upsilon_i +\upsilon_{i+2})  \sigma_{i+1}  \nonumber \\
&+& J_4 (\sigma_i  \upsilon_{i+2}+ \sigma_{i+2} \upsilon_i) + K_1  \sum_{i=1}^2 \sigma_i \sigma_{i+1}  \upsilon_i \upsilon_{i+1} + K_2   \sigma_i \sigma_{i+2}  \upsilon_i \upsilon_{i+2}  \nonumber \\
&+& K_3  \sigma_i \sigma_{i+1}  \sigma_{i+2} (\upsilon_i +\upsilon_{i+2}) + K_{3c}   \sigma_i \sigma_{i+1}  \sigma_{i+2} \upsilon_{i+1} +  K_{5}  \upsilon_i \upsilon_{i+1}  \upsilon_{i+2} (\sigma_i +\sigma_{i+2})  \nonumber \\
&+& K_{5c}  \upsilon_i \upsilon_{i+1}  \upsilon_{i+2}  \sigma_{i+1} + K_7 ( \sigma_i \sigma_{i+1} \upsilon_{i+1} \upsilon_{i+2} + \sigma_{i+1} \sigma_{i+2} \upsilon_i \upsilon_{i+1})  \nonumber \\
&+& K_8  \sigma_i \sigma_{i+1} \upsilon_i \upsilon_{i+2}   
+K_9  \sigma_i \sigma_{i+2} \upsilon_i \upsilon_{i+1} + R_6  \sigma_i \sigma_{i+1} \sigma_{i+2} \upsilon_i \upsilon_{i+1} \upsilon_{i+2}.
\end{eqnarray}
As seen the mentioned inequivalence leads to pairs of  effective couplings in place of a single one, and so  $(J_{p}, J_{pc})$ in place of original $j_p$; $(J_{3s}, J_{3v})$ in place of  $j_3$; $(K_3, K_{3c})$ in place of $k_3$; and  $(K_5, K_{5c})$ in place of $k_5$.
However, because the original sites are equivalent one has to make an arbitrary choice of the parameters and we assume: $j_p \rightarrow J_p,   j_3 \rightarrow  \frac{1}{2} (J_{3s}+J_{3v}),  k_3 \rightarrow K_3$ and $k_5 \rightarrow K_5$.



\begin{thebibliography}{}
\bibitem{Gal} G{\'a}lisov{\'a} L, Stre\u{c}ka J 2015 Physical Review E, {\bf  91},  0222134
\bibitem{Str}  Stre\u{c}ka J,  Al{'ee}cio R C, Lyra M L, and Rojas O 2016 J. Magn. Magn. Mater. {\bf  409}, 124
\bibitem{Souza}  de Souza S M, Rojas O 2017 Solid State Communications  {\bf 269}, 131
\bibitem{Rojas} Rojas O,  Stre\u{c}ka J,Lyra M L,  deSouza S M 2019 Physical Review E, {\bf 99}, 042117
\bibitem{Wei1}  Weiguo Yin, Frustration-driven unconventional phase transitions at finite temperature in a one-dimensional ladder Ising model 2020 arXiv:2006.08921v2.
\bibitem{Wei2}  Weiguo Yin, Finding and classifying an infinite number of cases of the practically perfect phase transition in an Ising model in one dimension 2020 arXiv:2006.15087v1.
\bibitem {Hut}Hutak T, Krokhmalskii T,  Rojas O, de Souza S M and Derzhko  O 2021  Phys. Lett. A  {\bf 387}, 127020
\bibitem{Tsv} Weichselbaum A, Weiguo Yin, Tsvelik A M 2021 Phys. Rev. B  {\bf 103}, 125120
\bibitem{Rojas2} Rojas O,  Valverde J S, de Souza S M 2009  Physica A {\bf 388}, 1419. 
\bibitem{Krok} Krokhmalskii T,  Hutak T, Rojas O,  de Souza S M, and  Derzhko O, Towards low-temperature peculiarities of thermodynamic quantities for decorated spin chains 2019 arXiv:1908.06419v1. 
\bibitem{MN} Nauenberg M, 1975 J. Math. Phys. {\bf 16}, 703.
\bibitem{Str2}  Strecka J,    An Introduction to the Ising model, Nova Science Publisher 2000 Chapter 4
                ISBN: 978-1-53618-145-6 
\bibitem{KW} Weron-Sznajd K,  Sznajd J 2005 Physica A {\bf 351} 593.
\bibitem{JS1} Sznajd J 2021 J. Stat. Mech.  013210.
\bibitem{Galam} Galam S 2008 Journal of Modern Physics C {\bf 19} 409-440 ; doi.org/10.1142/S0129183108012297.
\bibitem{KW1} Sznajd-Weron K 2002 Phys. Rev. E 66 046131.
\bibitem{KW2} Sznajd-Weron K 2004 Phys. Rev. E 70 037104.
\bibitem{Toral} Toral R,  Tessone C J 2007  Commun. Comput. Phys. {\bf 2}  177.
\bibitem{Hol}Ho\l yst J A, Kacperski K, Schweitzer F 2001 Annual Review of Computational Physics IX, World
Scientific, Singapore,  p 275 .
\bibitem{KW0} Weron-Sznajd K,  Sznajd J 2000 J. Modern Phys. C {\bf 11} 1157.  http://dx.doi.org/ 10.1142/S0129183100000936.

\end{thebibliography}
\end{document}